\documentclass[aps,amsfonts,nofootinbib]{revtex4}
\usepackage{epsfig}
\usepackage{graphicx}
\usepackage{amsmath}
\usepackage{amsbsy}
\usepackage{color}
\usepackage{epsfig}
\usepackage{graphicx}
\usepackage{amsmath}
\usepackage{amssymb}
\usepackage{bbm}
\usepackage{amsbsy}
\newcommand{\ee}{\end{equation}}
\newcommand{\bb}{\begin{equation}}
\newcommand{\eqb}{\begin{eqnarray}}
\newcommand{\eqf}{\end{eqnarray}}

\newcommand{\1}{{\'{\i}}}

\def\1{\'{\i}}
\begin{document}
\title{Particle-antiparticle asymmetry from magnetogenesis through the Landau mechanism}
\author{D. C\'arcamo}
\email{dacarcamo@uc.cl}
\affiliation{Departamento    de   F\'{\i}sica,    Pontificia   Universidad
  Cat\'olica de Chile, Casilla 306, Santiago 22, Chile}
  \author{Ashok Das}
\email{das@pas.rochester.edu}
\affiliation{Department of Physics and Astronomy, University of Rochester, Rochester, NY 14627-0171, USA
\\  and 
\\
 Saha Institute of Nuclear Physics, 1/AF Bidhannagar, Calcutta 700064, India}
\author{J. Gamboa }
\email{jgamboa55@gmail.com}
\affiliation{Departamento de  F\'{\i}sica, Universidad de  Santiago de
  Chile, Casilla 307, Santiago, Chile}
\author{M. Loewe}
\email{mloewe@fis.puc.cl}
\affiliation{Departamento    de   F\'{\i}sica,    Pontificia   Universidad
  Cat\'olica de Chile, Casilla 306, Santiago 22, Chile\\ and \\
  Centre for Theoretical Physics and Mathematical Physics, University of Cape Town, Rondebosch 7700, South Africa}

\begin{abstract}
Motivated by string theory an extension of the Landau problem to quantum field theory is considered.  We show that the commutator between momenta of the fields violates Lorentz and CPT invariance leading to an alternative method of understanding the question of particle-antiparticle asymmetry. The presence of magnetic field at very early moments of the universe would then suggest that the particle-antiparticle asymmetry can be understood as a consequence of magnetogenesis.
\end{abstract}

\maketitle

The quantum mechanics of a charged particle in a constant magnetic field is known as the Landau problem and is an interesting phenomenon. In this case, the Schr\"{o}dinger equation can be exactly solved and the energy levels as well as the degeneracy of states can be determined. The Landau problem with strong magnetic fields is currently a widely studied area in different fields such as solid state physics \cite{laugh}, astrophysics \cite{shapiro} and cosmology \cite{grasso,magne,gio}.  It also provides a beautiful formalism for implementing fractional statistics \cite{wilc} as well as a simple example of a quantum mechanical system with deformed commutators in the phase space \cite{nair}. The mechanical (kinematic) momenta do not commute in the presence of a magnetic field which is a nontrivial characteristic of the Landau problem and this has a profound theoretical \cite{laugh} and experimental impact \cite{von}. It is this noncommutativity of the momenta that we will like to explore in the following. 

In cosmology we know that there are various sources of magnetic fields including a primordial magnetic field. Therefore, in order to study physical phenomena at early stages of the universe, it would be of interest to  generalize the Landau problem to quantum field theories. The presence of a background magnetic field necessarily violates Lorentz invariance and will be reflected in any such theory. A direct generalization of the coupling of a field to a magnetic field can, of course, be done through the minimal coupling, but this would not lead to the non commutativity of momenta which is a characteristic of the Landau problem and which we are interested in.

In this work we propose a generalization of the Landau problem to quantum field theories motivated by the coupling of the string to an external magnetic field (in a target space sense). As we will show, this leads to the noncommutativity of momenta, a breaking of Lorentz invariance (and, therefore, CPT) as well as the particle-antiparticle symmetry. In this scenario, therefore, the primordial magnetic field becomes related to the particle-antiparticle symmetry breaking (or in some sense magneto genesis can become related to baryogenesis).

Let us recapitulate briefly the Landau problem, namely, the motion of a charged particle in the presence of a constant magnetic field $B$ along the $z$-axis. In this case ignoring the trivial motion along the $z$-axis, the dynamics in the plane is described by the Lagrangian in the symmetric gauge (summation over repeated indices is understood)
\begin{equation}
L = \frac{1}{2} (\dot{x}_{i})^{2} +  A_{i} \dot{x}_{i} = \frac{1}{2} (\dot{x}_{i})^{2} - \frac{B}{2} \epsilon_{ij} \dot{x}_{i} x_{j},\label{L1}
\end{equation}
where we have set the mass as well as the coupling ($e$) to unity and have chosen the symmetric gauge for the vector potential. Here $i=1,2$ and we have identified $x_{1}=x, x_{2}=y$ and the ``dot" denotes a derivative with respect to time. Because of the velocity dependent coupling, the canonical momentum is modified (from the free particle case)  and is given by
\begin{equation}
\Pi_{i} = \frac{\partial L}{\partial \dot{x}_{i}} = \dot{x}_{i} - \frac{B}{2} \epsilon_{ij} x_{j} = p_{i} - \frac{B}{2} \epsilon_{ij} x_{j},\label{M1}
\end{equation}
where $p_{i}$ denotes the mechanical (sometimes also known as the kinematic) momentum and is gauge invariant. The Hamiltonian for the system now follows to be
\begin{equation}
H = \Pi_{i} \dot{x}_{i} - L = \frac{1}{2} (p_{i})^{2},\label{H1}
\end{equation}
and the equal-time commutators between the basic variables can be determined to be ($\hbar = 1$)
\begin{equation}
[x_{i}, x_{j}] = 0,\quad [x_{i}, p_{j}] = i\delta_{ij},\quad [p_{i}, p_{j}] = iB \epsilon_{ij}.\label{C1}
\end{equation}
The Hamiltonian equations lead to
\begin{equation}
\dot{x}_{i} = p_{i},\quad \dot{p}_{i} = \ddot{x}_{i} = B \epsilon_{ij} p_{j} = B \epsilon_{ij} \dot{x}_{j},\label{E1}
\end{equation}
which coincides with the Euler-Lagrange equation following from \eqref{L1}. We note here that the deformation of the momentum commutator in the presence of a magnetic field in \eqref{C1} provides the simplest example of noncommutativity in quantum mechanics of recent interest (the non commutativity can also be transferred to the coordinate commutators through an appropriate change of basis).

In trying to generalize this phenomenon to field theory, as we have already mentioned, we will follow the approach from string theory in the sense that we will consider a magnetic field in the target space. Therefore, choosing a two component scalar field theory $\phi_{i} (t,x), i=1,2$ in two dimensions, the natural generalization of the Lagrangian in \eqref{L1} leads to a Lagrangian density ($c=1$)
\begin{equation}
{\cal L} = \frac{1}{2} \left((\dot{\phi}_{i})^{2} - (\phi'_{i})^{2}\right) - \frac{g_{2}}{2} \epsilon_{ij} \dot{\phi}_{i} \phi_{j},\label{L2}
\end{equation}
where a prime denotes a derivative with respect to $x$ and the constant $g_{2}$ plays the role of the magnetic field in \eqref{L1} (it has the dimension of energy). In this two dimensional field theory, however, there is a possibility to add a second quadratic term to the Lagrangian density in this generalization so that we can write (we assume $g_{1}, g_{2} > 0$)
\begin{equation}
{\cal L} =  \frac{1}{2} \left((\dot{\phi}_{i})^{2} - (\phi_{i}')^{2}\right) - \frac{g_{1}}{2} \dot{\phi}_{i} \phi_{i}' - \frac{g_{2}}{2} \epsilon_{ij} \dot{\phi}_{i} \phi_{j},\label{L3}
\end{equation}
where $g_{1}$ denotes a dimensionless parameter. Clearly, both the interaction terms violate Lorentz invariance. We will examine shortly the effects generated by $g_{1}$ and $g_{2}$ respectively.

It follows from the Lagrangian density in \eqref{L3} that the momenta canonically conjugate to the field variables $\phi_{i}$ are given by
\begin{equation}
\Pi_{i} (t,x) = \frac{\partial {\cal L}}{\partial \dot{\phi}_{i}} = \dot{\phi}_{i} - \frac{1}{2}\left(g_{1} \phi_{i}' + g_{2} \epsilon_{ij} \phi_{j}\right) = p_{i} - \frac{1}{2}\left(g_{1} \phi_{i}' + g_{2} \epsilon_{ij} \phi_{j}\right),\label{M2}
\end{equation}
which can be compared with \eqref{M1}. The Hamiltonian density now has the form
\begin{equation}
{\cal H} = \Pi_{i} \dot{\phi}_{i} - {\cal L} = \frac{1}{2}\left((\dot{\phi}_{i})^{2} + (\phi'_i)^{2}\right),\label{H2}
\end{equation}
as we would expect (see \eqref{H1}). The equal-time commutators between the basic variables can now be obtained to have the forms
\begin{equation}
[\phi_{i} (t,x),\phi_{j} (t,y)] = 0,\quad [\phi_{i} (t,x), p_{j} (t,y)] = i \delta_{ij} \delta (x-y),\quad [p_{i} (t,x), p_{j} (t,y)] = i {\cal F}_{ij} (x,y),\label{C2}
\end{equation}
where we have identified
\begin{equation}
{\cal F}_{ij} (x,y) = \left(g_{1} \delta_{ij} \partial_{x} + g_{2} \epsilon_{ij}\right) \delta (x-y).\label{F}
\end{equation}
It can be trivially checked that the structure in \eqref{F} satisfies the anti-symmetry as well as the Jacobi identity required of a (commutator) Hamiltonian structure. The Hamiltonian equations can now be derived which are also equivalent to the Euler-Lagrange equations following from \eqref{L3}, namely,
\begin{equation}
\ddot{\phi}_{i} - \phi_{i}'' - g_{1} \dot{\phi}_{i}' - g_{2} \epsilon_{ij} \dot{\phi}_{j} = 0.\label{E2}
\end{equation}

Using a plane wave solution of the form
\begin{equation}
\phi_{i}(t,x) = \phi_{i} (E,k)\, e^{-i (Et - kx)},
\end{equation}
equation \eqref{E2} takes the form
\begin{equation}
\left((-E^{2} + k^{2} - g_{1}EK) \delta_{ij} + ig_{2} E \epsilon_{ij}\right) \phi_{j} (E,k) = 0.
\end{equation}
The two coupled equations can be decoupled through a change of basis and take the forms
\begin{align}
& \left(E^{2} - (g_{2} - g_{1} k)E - k^{2}\right) \phi^{(+)} (E,k) = 0,\label{E3}\\
& \left(E^{2} + (g_{2} + g_{1} k)E - k^{2}\right) \phi^{(-)} (E,k) = 0,\label{E4}
\end{align}
where $\phi^{(\pm)}(E,k) = \phi_{1}(E,k) \pm i\phi_{2}(E,k)$ can be thought of as momentum space wave functions for particles and anti-particles respectively. The energy eigenvalues follow from \eqref{E3} and \eqref{E4} to correspond to
\begin{align}
E^{(+)}_{\pm} = \frac{1}{2}\left[- (g_{1}k - g_{2}) \pm \sqrt{(g_{1}k - g_{2})^{2} + 4k^{2}}\right],\label{soln1}\\
E^{(-)}_{\pm} = \frac{1}{2}\left[- (g_{1}k + g_{2}) \pm \sqrt{(g_{1}k + g_{2})^{2} + 4k^{2}}\right].\label{soln2}
\end{align}

We note that when $g_{2} = 0$, the two solutions in \eqref{soln1} and \eqref{soln2} coincide and lead to
\begin{equation}
E^{(+)}_{\pm} = E^{(-)}_{\pm} = E_{\pm} = \left[- \frac{g_{1}}{2}\, \epsilon (k) \pm \sqrt{1+\frac{g_{1}^{2}}{4}}\right] |k|,\label{soln3}
\end{equation}
where $\epsilon (k)$ denotes the sign function (namely, $+1$ for $k>0$ and $-1$ for $k< 0$).  It follows that the speed of propagation of these massless scalar excitations for $k << 1$ is
\bb 
\frac{E_\pm}{dk} = 1-\frac{g_1}{2} + {\cal O}(g_1^2),
\ee 
and therefore
\bb 
\delta v=1-v= \frac{g_1}{2}.  
\ee

However as $v =1$ is the maximum attainable speed, can extract bounds pair $g_1$ using data from the SME \cite{kost22}, and indeed one can see that $g_1$ coincides 
with $c_ {xt}$ in \cite{kost22} (see table I).

We see that, in this case,  the speed of scalar excitations is modified which is, of course, a consequence of the Lorentz violation by the interaction term with the coupling $g_{1}$. However, since $E^{(+)} - E^{(-)} = 0$, the term with the coupling $g_{1}$ does not lead to any particle-antiparticle asymmetry. 

\begin{table}[h]
\centerline{
\begin{tabular}{|c||c|c|c|c|c|c|}\hline
$\quad$ LIV tests $\quad$ & $\ \ \ \ \delta v_{\pm} \sim \ \ \ \ $ &\ \ \ \ Ref.\ \ \ \ \\
\hline \hline $\quad$Electron data $\quad$& $10^{-18}$ & \cite{kost22}
\\ \hline Proton data  & $10^{-20}$  & \cite{kost22} \\
\hline Pair creation processes  & $10^{-15}$ & \cite{altschul} \\
\hline Cold Atom Clock & $10^{-19}$ & \cite{wolf}\\
\hline
\end{tabular}}
\caption{Bounds on $\delta v$ given by different  tests. 
\label{lor}}
\end{table}

On the other hand, various tests of Lorentz invariance violation have already measured the possible changes in $c$ and the results are tabulated in table I. A bound on the strength of the coupling $g_{1}$ can, therefore, be obtained from these measurements and leads to $g_{1} < 10^{-16}$ from pair creations processes while proton data lead to 
$g_{1} < 10^{-20}$. Therefore, we conclude that the contributions coming from $g_{1}$ in the extra interaction term are negligible. Consequently, if we neglect $g_{1}$ in \eqref{soln1} and \eqref{soln2}, we obtain
\begin{align}
E_{\pm}^{(+)} & = \frac{g_{2}}{2} \pm \sqrt{k^{2} + \left(\frac{g_{2}}{2}\right)^{2}},\notag\\
E_{\pm}^{(-)} & = - \frac{g_{2}}{2} \pm \sqrt{k^{2} + \left(\frac{g_{2}}{2}\right)^{2}}.\label{dispersion}
\end{align}
We note that in this case $E^{(+)}_{\pm} \neq E^{(-)}_{\pm}$ and, in fact, the difference in energy between particles and antiparticles has the form
\begin{equation}
E^{(+)}_{\pm} - E^{(-)}_{\pm} = g_{2},\label{g2}
\end{equation}
so that we recognize $g_{2}$ as a measure of particle-antiparticle symmetry breaking. A nonzero value of the coupling $g_{2}$ necessarily breaks particle-antiparticle symmetry (in addition to Lorentz invariance) \cite{kost,jackiw,nos,jackiw1,bertolami,mewes,ko,AnSol} and this is the difference between the two Lorentz invariance violating terms in \eqref{L3}. Of course, as we have already pointed out $g_{2}$ can be thought of as a constant background magnetic field (in the target space sense) and, therefore, in this approach a background magnetic field can be thought of as a reason for particle-antiparticle symmetry breaking. 

If we generalize the dispersion relation \eqref{dispersion} to hold for fermions, it is easy to see that the parameter $g_{2}$ can even be related to the baryon asymmetry parameter in the following way. We note that the baryon asymmetry (particle-antiparticle asymmetry) is parameterized by the ratio \cite{bary}

\bb 
\eta = \frac{n-{\bar n}}{s} = \frac{n-{\bar n}}{7 n_{\gamma}}, \label{bbb}
\ee 
where $n, \bar{n}$ denote the densities for particles and antiparticles respectively,  $s$ is the entropy density which can be identified with $s= 7n_\gamma$ where $n_\gamma$ denotes the photon density. The difference $(n - \bar{n})$ can be calculated easily if we assume thermal equilibrium at the beginning of nucleosynthesis (around $kT = \beta^{-1} = 20$MeV), namely, the number difference between particles and antiparticles in 1 cm is obtained from equilibrium statistical mechanics to be
\begin{equation}
n - \bar{n} = \int_{-\infty}^{\infty}dk \left(\frac{1}{e^{\beta E_{+}^{(+)}} + 1} - \frac{1}{e^{\beta E_{+}^{(-)}} + 1}\right).
\end{equation}
Assuming that $g_{2}$ is small, we can Taylor expand the integrand to obtain the leading term 
\begin{equation}
n - \bar{n} = 2g_{2}\beta = \frac{g_{2}}{10 {\rm MeV-cm}}.
\end{equation} 
Substituting this into \eqref{bbb}, we obtain
\begin{equation}
\eta = \frac{g_{2}}{70 n_{\gamma} {\rm MeV-cm}},\label{relation}
\end{equation}
which relates $g_{2}$ to the baryon asymmetry parameter and the photon number density. On the other hand, as we have remarked earlier, $g_{2}$ can be thought of as a constant background magnetic field so that \eqref{relation} would relate baryogenesis with the presence of a constant background magnetic field which can possibly be the primordial magnetic field. Although our model is very simple and in $1+1$ dimensions, we can try to estimate $g_{2}$ by extrapolating the known observational data. For example, the asymmetry parameter is measured to be
\begin{equation}
\eta \sim 6 \times 10^{-10},
\end{equation}
and the number of photons in one cubic centimeter is known to be
\begin{equation}
n_{\gamma} \sim 400.
\end{equation}
If we naively reduce this to one dimension and take $n_{\gamma} \sim 7$/cm and plug in these numbers into \eqref{relation} we obtain
\begin{equation}
g_{2} \sim 3\times 10^{-1} {\rm eV}.\label{primordial}
\end{equation}
We are fully aware that it is improper to take observational data on the three dimensional universe and apply it to the predictions of a one dimensional model. Our idea is simply to illustrate how with a more realistic model in $3+1$ dimensions, such a mechanism can in fact lead to a connection between the primordial magnetic field and baryogenesis. In a realistic $3+1$ dimensional model, a direct bilinear fermion coupling to a magnetic field is not allowed because of arguments of renormalizability and possibly can arise through radiative corrections. As a result, the strength of such a coupling will be much smaller bringing down the estimate \eqref{primordial} significantly (besides other effects) and may in fact become comparable to the value of the primordial magnetic field.  Since in this scenario there are no transitions between baryons to antileptons and vice versa, the processes involving sphalerons do not seem to be relevant. However, this question is under study.

\medskip

\noindent\textbf{Acknowledgements}: This  work was supported by grants from
FONDECYT-Chile grant-1095106, 1095217, 1100777, 22110026 and Proyecto Anillos ACT119. One of us (J.G.) thanks to the Physics and Astronomy Department of the University of Rochester and Departamento de F\'{\i}sica of PUC for the hospitality.

\end{document}